\documentclass[reprint, superscriptaddress, amsmath, amssymb, aps,showpacs]{revtex4-1}

\usepackage{graphicx}
\usepackage{dcolumn}
\usepackage{bm}
\usepackage{mathtools}

\begin{document}

\title{Multiple resonant manipulation of qubits by train of pulses} 

\author{Gor A. Abovyan}
\email{gor.abovyan@ysu.am}
\affiliation{Yerevan State University, A. Manookyan 1, 0025, Yerevan, Armenia}
\affiliation{Institute for Physical Research, Ashtarak-2, 0203, Ashtarak, Armenia}

\author{Gagik Yu. Kryuchkyan}
\email{kryuchkyan@ysu.am}
\affiliation{Yerevan State University, Centre of Quantum Technologies and New Materials, Alex Manoogian 1, 0025, Yerevan, Armenia}
\affiliation{Institute for Physical Research, Ashtarak-2, 0203, Ashtarak, Armenia}

\date{\today}

\begin{abstract}
We present a systematic approach based on Bloch vector's treatment and the Magnus quantum electrodynamical formalism to study qubit manipulation by a train of pulses. These investigations include one of the  basic processes involved in quantum computation. The concrete calculations are performed for tunneling quantum dynamics, multiple resonance  and off-resonance excitations of qubit driven by Gaussian pulses. In this way, the populations of qubit states due to multiple resonant interactions are investigated for various operational regimes including: single-pulse excitation, two-pulse excitation with phase shift between pulse envelopes being controlling parameter and for excitation with sequential pulses.  In the last case, we  demonstrate the formation of quasienergetic  states and quasienergies of qubit  driven by train of identical pulses. In this case the transition probability of qubit exhibits aperiodic oscillations, but also becomes periodically regular for definite values of the quasienergy. 

\end{abstract}

\pacs{85.25.-j, 03.65.-w, 03.67.-a}
\maketitle


\section{\label{Introduction}Introduction}

Interactions of quantum systems with pulse train are of great interest for many scientific and technological applications, such as qubit manipulation or quantum computing, secure communications and ultra-precise measurements. This direction is  similar to the method of nuclear magnetic resonance for  the radio-frequency control and manipulation of spin systems and atoms 
\cite{Hahn,Haeb}. In this way, the train of pulses with the name of composite pulses has been originally developed in a pioneering work \cite{Levitt}.

The method of control of quantum systems by pulse train has also been applied in atomic physics (see, for example: \cite{Vitanov}, \cite{Toros}), in magnetometry with solid state quantum sensors \cite{Aiello} and molecular spectroscopy  \cite{Scherer}. Some other applications have been noted in Ref. \cite{Kyos}.

This direction also involves the tunneling dynamics of time-dependently driven nonlinear quantum systems. The physics of driven quantum tunneling are associated with a wide variety of interesting phenomena and effects \cite{Grifoni}.

On the other hand, decoupling methods based on the multiple pulse techniques are very perspective for open quantum systems, for decoherence control in quantum engineering \cite{Viola,Viola1,Khodj}.
In this way, experimental applications have been obtained for trapped ions \cite{Biercuk,Bonato,Uys}, atomic ensembles \cite{Sagi} and  spin-based devices \cite{Du}.

In this spirit, we emphasize the idea of improving the degree of quantum effects in open systems, 
as well as obtaining qualitatively new quantum effects through application of the sequence of tailored pulses. This approach was recently exploited for formation of high degree continuous-variable entanglement in the nondegenerate optical parametric oscillator \cite{Adam,Manuk,Adam1,Adam2}, for investigation of quantum interference in mesoscopic domain \cite{Gevorg}, for production of Fock states and for realizing photon blockade in Kerr nonlinear resonator driven by pulse train \cite{Faraon,Gevorg1,Gevorg2,Hovsepyan}, and for demonstration of quantum chaos at level of few quanta \cite{Gevorg3,Chew}.

The control of qubits by pulses is a basic process involved in quantum computation. In this direction superconducting qubits are among the most promising systems for the realization of quantum computation. Recently, Rabi oscillations, manipulation and control of states of single superconducting qubit by resonant microwave pulses have been extensively studied \cite{Naka,Marti,Chio,Vion,Koch}.

Recently, the dynamics of superconducting qubit driven by external field with time-modulated amplitude and the phenomenon of Rabi oscillations have been considered \cite{Abov} close to the experimental scheme on the frequency-modulated transmon qubit \cite{Jian}. Superconducting qubits usually have short coherence time, therefore to decrease the time for performing gate operations a large-amplitude external fields should be applied. The dynamics of a qubit driven by large-amplitude external fields in the case of driving around the region of avoided level crossing has been also studied (see, \cite{Ashhab,Shevchenko} for reviews).

In the present paper we present a systematic approach for investigation of tunneling quantum dynamics  of two-level systems, particularly, superconducting qubits interacting with a single pulse as well as with a train of pulses. In this way, we analyse the Rabi model on base of both the Bloch vector's treatment and the Magnus QED formalism for calculation of time-evolution operator. The Bloch vector, three-dimensional unit vector, provides a convenient pictorial representation of qubit state \cite{Nielsen} while the Magnus formalism \cite{Blanes,BlanesPhysRep} leads to expansion of the time-evolution operator without well known Dyson time-ordering in QED.  We formulate the Magnus formalism in so-called Furry representation that allows to apply the  rotating-wave approximation (RWA) as well as to consider naturally the system beyond RWA. In addition, we  demonstrate that this approach is also valid for investigation of pulse-driven  multi-qubit systems.

This approach allows to obtain general  explicit   solutions for single-qubit populations in dependence
from time-dependent Rabi frequency that can be calculated in details on base of the Magnus expansion.
Note that this result cannot been obtained in the framework of Dyson time-dependent perturbation theory and is valid for various interactions including also the regime beyond rotating wave approximation. On the other side, using the  Magnus formalism in the Furry representation allows us to formulate an effective truncation procedure for Magnus series. So it will be strongy shown that in RWA the operator of time-evolution is only determined by the first term of Magnus series and effects beyond RWA can be calculated as corrections to the resonant part through the high-order Magnus terms.

It is demonstrated in this paper that the results can be also obtained for a wide range of frequencies of driving field corresponding to multiple  resonant excitation regimes of qubit on one side and involve  interactions with both  single pulse or the train of pulses with arbitrary envelopes on the other side.
The concrete calculations will be done for the train of pulses with Gaussian  envelopes and  with different  phases. The duration of pulses, time intervals between them and the relative phases of pulses are the free  control parameters.  Thus, this configuration  provides more flexibility for manipulation and control of qubit states, particularly, it allows us to engineer the  phase-controllable populations of qubit states. Such approach is useful for performing  a large number of quantum gates that usually are designed with atom like qubits interacting with applied driving pulses with finite durations. 

Another interesting regime  occurs when the pulse train consists of identical pulses and hence displays the periodicity in time. We will demonstrate that  quasienergies and quasienergetic states (QES) (or so-called the Floquet states) for the qubit driven by train of identical pulses are realized in multiple resonant regimes. Thus, the regular oscillations  of the state population is realized if the factor of the quasienergy and the period becomes equal to $m\pi$, where $m=1, 2,...$.

Note, that at first, the QES of the composite system consisting of an atom and time-periodic e.m. field have been considered first time in the papers  \cite{Shirley,Zeldovitch,*ZeldovitchZh,Ritus,*RitusZh}. These states provide a classical counterpart to well known atomic-dressed states \cite{Cohen} in which the coupling to the laser is described by a classical field, whereas the  coupling to the vacuum must be described in second quantization.

There have been several experiments on nonlinear and quantum optics that have been interpreted in terms of quasienergy levels including basic experiments on the resonance fluorescence and the probe absorption spectroscopy for a two-level atom in a strong laser field. 
QESs for a two-level atom in the bichromatic field have also been studied in a series of papers (see, for example, \cite{Freedhoff,*Freedhoff1,Jakob,Jakob1,Jakob2}. Applications of QES and quasienergies to Josephson qubits in driving field have been done in several papers \cite{Shevchenko,Russomanno,Silveri,Tuorila}, including review paper on Landau-Zener-St{\"u}ckelberg interferometry \cite{Shevchenko}, probe spectroscopy of QES \cite{Silveri}, application of the Floquet theory to Cooper pair pumping \cite{Russomanno}, and observation of Stark effect and generalized Bloch-Siegert shift in the experiment with  superconducting qubit probed by resonant absorption via a cavity \cite{Tuorila}. The experiments on the Rabi oscillations in monochromatically driven Josephson qubits have been performed and interpreted on the base of dressed states \cite{Johan,Naka}. 

The paper is organized as follows. The single driven qubit with time-dependent  interaction Hamiltonians along  the $z$ axis is studied in Sec. \ref{PDSQ} in the Bloch vector's treatment and by using the Magnus expansion. In this section,  the general results for the  populations of qubit states are reported, particularly,  for the regimes of  multiple resonant excitations including: single-pulse excitation and two-pulse excitation with phase shift between pulse envelopes being controlling parameter. Moreover, the modifications of this approach for multi-qubit systems, including details for two-qubit system and off-resonance interaction beyond RWA are presented in Sec. \ref{PDSQ}. The multiple resonant excitations with sequential pulses are considered in Sec. \ref{MRESP}. This analysis involve also manipulation of qubit by train of identical pulses as well as formation of QES and quasienergies. Finally, some conclusions are given in Sec. \ref{conclusion}.

\section{Pulsed dynamics of a single qubit in the Furry picture and in the Bloch vector's path}\label{PDSQ}

In this section,  we derive general expressions for populations of a single qubit as well as we consider the  application of the approach for multi-qubits systems, particularly, for coupled two qubits. However, detailed calculations  will be done for a single driven qubit  for various multiple resonant excitation regimes including: single-pulse excitation and  two-pulse excitation with phase shift between pulse envelopes being controlling parameter.

\subsection{Time-evolution of qubits on base of the Magnus expansion. The probability of excitation and the Rabi frequency}

At first, we consider the qubit driven by a field of pulses with arbitrary envelopes on base of the Hamiltonian
\begin{equation}\label{Ham1}
\hat{H}(t)=-g(t)\hat{\sigma}_{z}+\Delta\hat{\sigma}_{x},
\end{equation}
where  time-dependent  interaction takes place along  the $z$ axis with the coupling constant
\begin{equation}\label{SigmaZFactor}
g(t)=\frac{1}{2}\varepsilon_{0}+A(t)\cos(\omega t+\theta).
\end{equation}

Here $\varepsilon_{0}$ is the electronic energy difference between the ground and excited states of the qubit  in the absence of the driving field,  $\Delta$  is the tunneling amplitude between the basis states, $g(t)$ includes the interaction between the external time-dependent field and the two-level system, and  $A(t)$ is the amplitude of external field, $\hat{\sigma}_{x}$,  $\hat{\sigma}_{y}$, $\hat{\sigma}_{z}$ are three usual Pauli matrices. The Hamiltonian (\ref{Ham1}) describes various physical systems  including quantum two-level systems with nonlinear tunneling mechanism driven by external field \cite{Grifoni}  and also superconducting qubit or JJ artificial atoms with $\Delta$ being the Josephson coupling energy. Up to now,  numerous proposals and demonstrations of quantum phenomena in superconducting qubit have been demonstrated in  the fields of atomic physics and quantum optics (see, for example review papers \cite{Clarke,You}).

We start with time-evolution equation of the system
\begin{equation}
i\frac{\partial}{\partial t}|\Psi(t)\rangle=\hat{H}(t)|\Psi(t)\rangle
\end{equation}
in the Farry picture $ |\Psi(t)\rangle=\hat{U}|\Phi(t)\rangle$. The transformation leads to the equations 
\begin{eqnarray}
i\frac{\partial\hat{U}}{\partial t}=-g(t)\hat{\sigma}_{z}\hat{U},\\
i\frac{\partial}{\partial t}|\Phi\rangle=\Delta\hat{U}^{-1}\hat{\sigma}_{x}\hat{U}|\Phi\rangle.
\end{eqnarray}
The operator  $\hat{U}$ is calculated as
\begin{equation}
\hat{U}=\exp\left(i\int_{0}^{t}g(t')dt'\cdot\hat{\sigma}_{z}\right)
\end{equation}
and time-evolution equation of the system can be rewritten in the following form 
\begin{equation}
i\frac{\partial}{\partial t}|\Phi(t)\rangle=\hat{h}(t)|\Phi(t)\rangle,
\end{equation}
where 
\begin{equation}\label{FarryHam}
\hat{h}=\Delta\left[e^{-2i\Lambda(t)}\hat{\sigma}_{+}+e^{2i\Lambda(t)}\hat{\sigma}_{-}\right],
\end{equation}
where $\hat{\sigma}_{+}=(\hat{\sigma}_{x}+i\hat{\sigma}_{y})/2$, $\hat{\sigma}_{-}=(\hat{\sigma}_{x}-i\hat{\sigma}_{y})/2$ and
\begin{equation}\label{lambda}
\Lambda(t)=\int_{0}^{t}g(t')dt'.
\end{equation}

Following a standard perturbation approach  time-evolution of quantum states is described in the Dyson expansion form as
\begin{equation}
|\Phi(t)\rangle=T\exp\left(-i\int_{0}^{t}\hat{h}(t')dt'\right)|\Phi(0)\rangle,
\end{equation}
where $T$ is the time ordering operator and $|\Phi(0)\rangle$ is an initial state of the system. 
Considering  below high-order terms in the perturbation expansion we note that the
Hamiltonian does not commute at different times due to Dyson-type time dependent integrations.  This circumstance usually makes the analysis of high-order processes complicated. On the other hand the unitary evaluation operator $\hat{S}(t,0)$ can be also presented in the other form: so-called Magnus expansion (see, for example \cite{Blanes}), which seems to be more preferable for some specific systems than the standard Dyson expansion. In this representation the time-evolution operator is a true exponential operator
\begin{equation}
\hat{S}(t,0)=T\exp\left(-i\int_{0}^{t}\hat{h}(t')dt'\right)=\exp(\hat{\Omega}(t)),
\end{equation}
where $\hat{\Omega}(t=0)=0$ and a series Magnus expansion for the operator in the exponent takes place
\begin{equation}\label{expansion}
\hat{\Omega}(t)=\sum^{\infty}_{k=1}{\hat{\Omega}_{k}(t)}.
\end{equation}
For instance, the three terms of the series are given as
\begin{eqnarray}\label{MagnusEx}
\hat{\Omega}_{1}(t) & = & -i\int_{0}^{t}\hat{h}(t')dt',\nonumber\\
\hat{\Omega}_{2}(t) & = & -\frac{1}{2}\int_{0}^{t}dt'\int_{0}^{t'}dt''[\hat{h}(t'), \hat{h}(t'')],\\
\hat{\Omega}_{3}(t) & = & \frac{i}{6}\int_{0}^{t}d t_{1}\int_{0}^{t_{1}}d t_{2}\int_{0}^{t_{2}}d t_{3}
\bigg([\hat{h}(t_{1}),[\hat{h}(t_{2}),\hat{h}(t_{3})]]\nonumber\\
& & +[\hat{h}(t_{3}),[\hat{h}(t_{2}),\hat{h}(t_{1})]]\bigg)\nonumber.
\end{eqnarray}

Thus, the Magnus approach provides the  approximate exponential representations of the time-evolution operator of the system.  It is not difficult to illustrate  a connection between Magnus series and Dyson time-dependent perturbative series. For instance, it is easy to check that the second-order Dyson term of perturbation theory corresponds to the second-order Magnus expansion. It follows from the following equality
\begin{eqnarray}
\int^{t}_{t_{0}}dt_{1}\int^{t}_{t_{0}}dt_{2}T\left(\hat{h}(t_{1})\hat{h}(t_{2})\right)=\left(\int^{t}_{t_{0}}dt_{1}\hat{h}(t_{1})\right)^{2}\nonumber\\
+\int^{t}_{t_{0}}dt_{1}\int^{t_{1}}_{t_{0}}dt_{2}[\hat{h}(t_{1}),\hat{h}(t_{2})]
\end{eqnarray}
and the structure of  $\hat{\Omega}_{1}(t)$ and  $\hat{\Omega}_{2}(t)$.

A simple description of a two-state system presents the state vector as a point on a two-dimensional sphere, the Bloch sphere. Therefore, we  represent the evolution operator as an infinitesimal rotation of the Bloch vector in abstract space
\begin{equation}
|\Phi(t+dt)\rangle=\hat{S}(t+dt,t)|\Phi(t)\rangle,
\end{equation}

\begin{equation}
\hat{S}(t+dt,t)=\hat{I}-i\frac{d\theta(t)}{2}(\vec{n}\hat{\vec{\sigma}})
\end{equation}
through an angle $d\theta=2\Delta dt$
about the axis directed along the unit vector with the components 
\begin{eqnarray}
n_{x} & = & \cos(2\Lambda(t)),\nonumber\\
n_{y} & = & \sin(2\Lambda(t)),\\
n_{z} & = & 0.\nonumber
\end{eqnarray}

Note, that interpretation of the Bloch vector's path beyond rotating wave approximation for a two-level system interacting with  a monochromatic field has been done recently \cite{Benenti}.

It is easy to realize that in general the time-evolution operator in the path of Bloch ball can be represented in the simple form for all terms of the series expansion (\ref{expansion}). Indeed, the all terms of the Magnus expansion are only expressed through the Pauli matrices $\hat{\sigma}_{x}$, $\hat{\sigma}_{x}$, $\hat{\sigma}_{x}$ and hence the following representation of time-evolution operator can be derived
\begin{multline}\label{transPauliMatrix}
\hat{S}(t,0)=\exp(\hat{\Omega}(t))=\exp\left(-i\vec{G}(t)\hat{\vec{\sigma}}\right)\\
=\hat{I}\cos(|\vec{G}(t)|)-i(\vec{\rho}(t)\hat{\vec{\sigma}})\sin(|\vec{G}(t)|),
\end{multline}
where the vector $\vec{\rho}(t)=\vec{G}(t)/|\vec{G}(t)|$. 
In this way, the problem is reduced to calculation of the function $|\vec{G}(t)|$ and  the vector $\vec{\rho}$  as the series of Magnus expansion. In the second-order Magnus expansion we obtain from Eqs. (\ref{MagnusEx})
\begin{eqnarray}\label{phaseGComp}
G_{x}(t) & = & \Delta\int_{0}^{t}dt'\cos(2\Lambda(t')),\nonumber\\
G_{y}(t) & = & \Delta\int_{0}^{t}dt'\sin(2\Lambda(t')),\\
G_{z}(t) & = & \Delta^{2}\int_{0}^{t}dt'\int_{0}^{t'}dt''\sin\left[2(\Lambda(t'')-\Lambda(t'))\right],\nonumber
\end{eqnarray}
and $G(t)=|\vec{G}(t)|=\sqrt{G_{x}^{2}(t)+G_{y}^{2}(t)+G_{z}^{2}(t)}$.
Thus, the vector state reads as
\begin{equation}\label{stateEqMag}
|\Phi(t)\rangle=\exp\left(-i\vec{G}(t)\hat{\vec{\sigma}}\right)|\Phi(0)\rangle
\end{equation}
and hence
\begin{equation}
|\Psi(t)\rangle=\hat{U}(t)\exp\left(-i\vec{G}(t)\hat{\vec{\sigma}}\right)|\Psi(0)\rangle.
\end{equation}

On the whole for the probability amplitudes of qubit states
\begin{equation}\label{stateVector}
|\Psi(t)\rangle=C_{1}(t)|1\rangle+C_{2}(t)|2\rangle
\end{equation}
 we obtain
 \begin{subequations}\label{Cs_expl}
\begin{eqnarray}
C_{1}(t) & = & e^{i\Lambda(t)}[\cos(G(t))-i\rho_{z}\sin(G(t))]C_{1}(0)\nonumber\\
& + & e^{i\Lambda(t)}(-i\rho_{x}-\rho_{y})\sin(G(t))C_{2}(0),\\
C_{2}(t) & = & e^{-i\Lambda(t)}(-i\rho_{x}+\rho_{y})\sin(G(t))C_{1}(0)\nonumber\\
& + & e^{-i\Lambda(t)}[\cos(G(t))+i\rho_{z}\sin(G(t))]C_{2}(0).
\end{eqnarray}
\end{subequations}
For the system that initially was in the ground  state $|1\rangle$, $C_{1}(0)=1$, $C_{2}(0)=0$, the population distribution $P_{2}(t)$ of the state $|2\rangle$ is
\begin{equation}\label{P2}
P_{2}(t)=(1-\rho_{z}^{2})\sin^{2}(G(t)).
\end{equation}

We notice that this result is obtained in the most general form for arbitrary time-dependent field amplitude $A(t)$ and for a wide range of external field frequencies.  In the general case, however, there is not compact expressions for the components of the function $\vec{G}(t)$ that are given  as the  Magnus series expansions. The first term of this expansion (\ref{expansion}) coincides exactly with the simple exponent  solution of time-evolution operator with the Hamiltonian $\hat{h}(t)$. The every other $n$-th term $\hat{\Omega}_{n}(t)$ contains multiple integral of combinations of $n-1$ commutators containing $n$ Hamiltonian operators $\hat{h}(t)$.  Nevertheless, the result (\ref{P2}) has some advantages from a computational point of view. Particularly, it is easy to get the simplest expressions for the population of qubit  in RWA describing interaction of  two-level system and a single electromagnetic  mode for the case when the frequency of mode is near resonance with  the qubit transition frequency and the coupling constant of interaction is weak (for atomic systems, see \cite{Allen,Shore,Meystre}).

\subsection{Multiple resonant excitations}

In this subsection the case of multiple resonant interaction with  pulses is considered. The resonance condition is formulated using the requirement that the oscillating terms in time are vanished in the integrals of Eqs. (\ref{phaseGComp}).  Thus, this condition is formulated for the  frequency $\omega$  and the electronic energy difference as $\varepsilon_{0}-N\omega=\Delta_{N}\ll\varepsilon_{0}$. At first, we consider Eqs. (\ref{SigmaZFactor}), (\ref{lambda}) in the following form
\begin{multline}\label{lambda1}
\Lambda(t)=\frac{1}{2}\varepsilon_{0}t+\int_{0}^{t}A(t')\cos(\omega t'+\theta)dt'\\
=\frac{1}{2}\varepsilon_{0}t+A(t)\frac{\sin(\omega t+\theta)}{\omega}-\int_{0}^{t}\frac{\sin(\omega t'+\theta)}{\omega}\frac{\partial A(t')}{\partial t'}dt'
\end{multline}
for the case of adiabatic pulses for which duration of pulses $T\gg\frac{2\pi}{\omega}$.  In this approximation the expression (\ref{lambda1}) is simplified as 
\begin{equation}\label{lambdaApprox}
\Lambda(t)=\frac{1}{2}\varepsilon_{0}t+A(t)\frac{\sin(\omega t+\theta)}{\omega}.
\end{equation}
Indeed, the last integral in Eq. (\ref{lambda1}) can be omitted in comparison with the other two terms in the adiabatic regime, that it is easy to check for pulses with Gaussian envelope
\begin{equation}\label{GaussEnvelope}
A_{k}(t)=A_{0}e^{-\frac{(t-t_{k})^{2}}{T^{2}}}.
\end{equation}

Calculating Eqs. (\ref{phaseGComp}) with the approximated formula (\ref{lambdaApprox}) we also  use the well known formulae with the Bessel functions. In the result we can obtain
\begin{equation}
e^{i\left(\varepsilon_{0}t+2A\frac{\sin(\omega t+\theta)}{\omega}\right)}=e^{i\varepsilon_{0}t}\sum_{n}J_{n}\left(\frac{2A(t)}{\omega}\right)e^{in(\omega t+\theta)},
\end{equation}
where $J_{n}(x)$ is $n$-th order Bessel function of the first kind. 
For the case of exact $N$-th order resonance, $\varepsilon_{0}=N\omega$, we get
\begin{multline}
e^{2i\Lambda(t)}\approx J_{-N}\left(\frac{2A(t)}{\omega}\right)e^{-iN\theta}\\
=(-1)^{N}J_{N}\left(\frac{2A(t)}{\omega}\right)e^{-iN\theta}.
\end{multline}

In this approximation the Hamiltonian (\ref{FarryHam}) transforms to
\begin{equation}\label{singlePulseHam}
\hat{h}(t)=(-1)^{N}\Delta J_{N}\left(\frac{2A(t)}{\omega}\right)\left[\hat{\sigma}_{+}e^{iN\theta}+\hat{\sigma}_{-}e^{-iN\theta}\right].
\end{equation}
Thus, the components of the vector $\vec{G}(t)$ up to second-order of Magnus expansion are calculated as
\begin{eqnarray}\label{ApproxG}
G_{x}(t) & \approx & (-1)^{N}\Delta\cos(N\theta)\int_{0}^{t}J_{N}\left(\frac{2A(t')}{\omega}\right)dt',\nonumber\\
G_{y}(t) & \approx & (-1)^{N}\Delta\sin(N\theta)\int_{0}^{t}J_{N}\left(\frac{2A(t')}{\omega}\right)dt',\\
G_{z}(t) & \approx & 0.\nonumber
\end{eqnarray}
As we see, these Eqs. (\ref{ApproxG}) don't involve the terms of second-order as has been noted in Introduction.  The   component $G_{z}(t)$ going from the second-order Magnus term and hence equal to zero in RWA approximation. It is easy to explain this truncation of Magnus series in RWA from the general point of view. Really, the high-order terms of Magnus expansion $ \hat{\Omega}_{2}(t)$, $\hat{\Omega}_{3}(t)$, ...  involve commutators of Hamiltonian operators  $[\hat{h}(t_{1}), \hat{h}(t_{2})]$, $ [\hat{h}(t_{1}),[\hat{h}(t_{2}),\hat{h}(t_{3})]]$, ... that are equal to zero for the case of  Hamiltonian in RWA (see, formula (30)). Thus, in RWA the operator of time-evolution is determined only by lowest exponential term involved $\hat{h}(t)$.

On the whole for the population of excited state we obtain the following result
\begin{equation}\label{multipleP2}
P_{2}(t)=\sin^{2}\left(\Delta\int_{0}^{t}J_{N}\left(\frac{2A(t')}{\omega}\right)dt'\right).
\end{equation}

\begin{figure}
 \includegraphics[height=4.5cm]{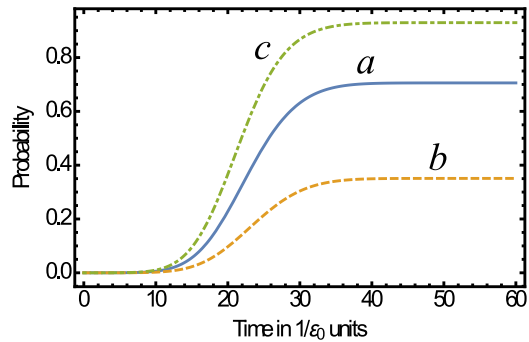}
\caption{\label{singlePulse1} The population of excited state $P_{2}(t)$  for the case of first-order resonance in dependence of time-intervals  in units of $1/\epsilon_{0}$. The parameters are:  $\Delta/\varepsilon_{0} = 0.3$, $T\varepsilon_{0}=10$,     $\omega/\varepsilon_{0} = 1$.  Solid line (a), $A_{0}/\varepsilon_{0}=0.19$;   dashed line (b),  $A_{0}/\varepsilon_{0}=0.12$;  dotdashed line (c), $A_{0}/\varepsilon_{0}=0.25$.}
\end{figure}

\begin{figure}
 \includegraphics[height=4.5cm]{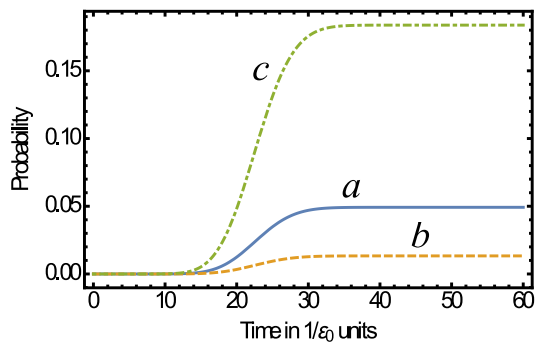}
\caption{\label{singlePulse2}The population of excited state $P_{2}(t)$ for the case of second-order resonance  in dependence of time-intervals  in units of $1/\epsilon_{0}$. The parameters are:  $\Delta/\varepsilon_{0} = 0.3$,  $T\varepsilon_{0}=10$,     $\omega/\varepsilon_{0} = 1$.  Solid line (a), $A_{0}/\varepsilon_{0}=0.35$; dashed line (b),  $A_{0}/\varepsilon_{0}=0.25$;  dotdashed line (c), $A_{0}/\varepsilon_{0}=0.5$.}
\end{figure}

Below we present application of this result for excitation of qubit by a single pulse at the frequency $\omega$ and with Gaussian envelope (see, Eq. (\ref{GaussEnvelope})), if $C_{1}(0) = 1$, $C_{2}(0) = 0$.
The results corresponding to the case of first-order resonance are depicted in Fig. \ref{singlePulse1}. As we see, the probability allows for a region of high transition probability with increasing the amplitude of single pulse.  The case of second-order resonance is demonstrated in  Fig. \ref{singlePulse2}.  As we see,  in this case the transition probability $P_{2}(t)$  is small in opposite to the  first-order resonance configuration. As will be demonstrated below the complete population inversion takes place for both resonant  configurations in cases of multi-pulse excitations.

\subsection{Qubit excitation beyond RWA: calculations up to the second-order of Magnus expansion}

\begin{figure}
 \includegraphics[width=7cm]{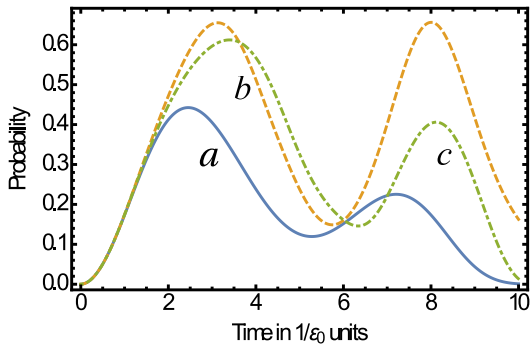}
\caption{\label{no_rwa}The population of excited state $P_{2}(t)$ beyond RWA in dependence of time-intervals in units of $1/\varepsilon_{0}$ for $C_{1}(0) = 1$, $C_{2}(0) = 0$ and the parameters: $\Delta/\varepsilon_{0} = 0.45$, $A_{0}/\varepsilon_{0}=0.4$,  $T\varepsilon_{0}=3.5$. Solid line (a), $\omega/\varepsilon_{0} = 0.5$; dashed line (b), $\omega/\varepsilon_{0} = 1.5$; dotdashed line (c), $\omega/\varepsilon_{0} = 1.1$.}
\end{figure}

In this subsection we shortly discuss above introduced approach in the regime beyond RWA, i.e. for the Rabi pulsed model in the
regimes where the RWA can not be applied. In this way, we consider single-qubit off-resonance interaction with strong external  Gaussian pulses. We calculate numerically qubit populations due to non-resonant excitation on the general formulas (\ref{Cs_expl}), (\ref{P2}) with high-order Magnus terms. As it can be seen from Eqs. (\ref{phaseGComp}),  the  first-order Magnus terms (proportional to  $\Delta$) are $G_{x}(t)$ and  $G_{y}(t) $ components of the vector $\vec{G}(t)$, while the second-order term (proportional to $\Delta^{2}$) involves only $z$ component, $G_{z}(t)$.  On the whole, the  Magnus expansion up to the  second-order involve all component  of the Bloch vector in Eq.(\ref{P2}) and third-order and high-order terms only are modified  these components. As shows calculations the small parameter of this expansion is the ratio $\Delta/\varepsilon_{0}<1$. Thus, below  we study numerically the temporal evolution of the qubit state vector (\ref{stateVector}), (\ref{Cs_expl}) up to the second-order Magnus terms. Note, that in calculations based on the formulas (\ref{phaseGComp}) the terms depending on $\omega+\varepsilon_{0}$ oscillate very rapidly
and have been  neglected in the rotating wave approximation (see, subsection A).
In this subsection, we will explore the effects of these counter-rotating terms
beyond RWA on calculation on the formulas (\ref{phaseGComp}) including these oscillation terms. The typical results are depicted in Fig.\ref{no_rwa} for three cases of non-resonance interaction with the frequencies of the external field: $\omega/\varepsilon_{0}=0.5$,  $\omega/\varepsilon_{0}=1.1$ and $\omega/\varepsilon_{0}=1.5$.  As we see, in this approach beyond RWA the population displays time-dependent  oscillations due to involving counter-rotating terms in qubit dynamics in contrast to analogous results for the resonant case depicted in Fig.\ref{singlePulse1}. The other peculiarity with respect to the resonance case is  that population of qubit decreases in time during the pulse. This effect is stipulated by the negative term $\rho_{z}^{2}$ in Eq. (\ref{P2}) appearing in the second-order Magnus expansion. The Fig.\ref{no_rwa} also shows an asymmetry  
in non-resonant excitation of qubit state connected with values of the frequency relatively to the energy difference between  ground and excited states of the qubit. Indeed, the population for the case of $\omega/\varepsilon_{0}=1.5$ exceeds analogous result for $\omega/\varepsilon_{0}=0.5$. It should be mentioned that  calculation of the third-order Magnus term (see, Eqs. (\ref{MagnusEx}))  for used parameters is only slightly change the obtained results.

\subsection{Phase-dependent two-pulse resonant excitation}

In this subsection the dynamic of qubit in the result of  two-pulse excitation is considered. We assume that the external field consists of two pulses with the same frequencies but with different amplitudes,
$A(t)=A_{1}(t)+A_{2}(t)$. The exponential function in this case reads as
\begin{equation}
\Lambda(t)=\frac{1}{2}\varepsilon_{0}t+A_{1}(t)\frac{\sin(\omega t+\theta_{1})}{\omega}+A_{2}(t)\frac{\sin(\omega t+\theta_{2})}{\omega}
\end{equation}
and the following expression can be obtained
\begin{eqnarray}
e^{2i\Lambda(t)}=\sum_{n_{1},n_{2}}J_{n_{1}}\left(\frac{2A_{1}(t)}{\omega}\right)J_{n_{2}}\left(\frac{2A_{2}(t)}{\omega}\right)\times\nonumber\\
e^{i\left[\epsilon_{0}+(n_{1}+n_{2})\omega\right]t}e^{in_{1}\theta_{1}}e^{in_{2}\theta_{2}}.
\end{eqnarray}

We assume  multiple resonant interaction  if  $\varepsilon_{0}-N\omega=0$, 
where $n_{1}+n_{2}=-N$. In RWA we reach
\begin{eqnarray}\label{exResPart}
e^{2i\Lambda(t)}=(-1)^{N}e^{-iN\theta_{2}}J_{N}(w(t))\left(\frac{Z-ze^{-i\phi}}{Z-ze^{i\phi}}\right)^{\frac{N}{2}},
\end{eqnarray}
where $w(t)=(Z^{2}+z^{2}-2zZ\cos(\phi))^{1/2}$, $z=\frac{2A_{1}(t)}{\omega}$, $Z=\frac{2A_{2}(t)}{\omega}$, $\phi=\theta_{1}-\theta_{2}+\pi$. This formula is valid when $\left|ze^{\pm i\phi}\right|<|Z|$.

This formula can be applied for time evolution that involves independent interactions with pulses, i.e. there is time interval between two interactions subjected by pulses, separated by interaction-free interval. Each of the independent interactions are described by a slowly varying RWA Hamiltonian (\ref{singlePulseHam}). Nevertheless, we assume that  pulses depend on the phases of the field envelopes. 
Thus, in this regime the complete expression for exponential function for an arbitrary moment in time reads as 
\begin{multline}\label{expExpand}
e^{2i\Lambda(t)}=(-1)^{N}\bigg[J_{N}\left(\frac{2A_{1}(t)}{\omega}\right)e^{-iN\theta_{1}}\\
+J_{N}\left(\frac{2A_{2}(t)}{\omega}\right)e^{-iN\theta_{2}}\bigg].
\end{multline}   
On the whole  the population of excited state is calculated from (\ref{phaseGComp}), (\ref{P2}) and (\ref{expExpand}) as
\begin{equation}
P_{2}(t)=\sin^{2}\left(\Delta\sqrt{w(t)}\right),
\end{equation}
where
\begin{equation}
w(t)=j_{1}^{2}(t)+j_{2}^{2}(t)+2\cos(N(\theta_{2}-\theta_{1}))j_{1}(t)j_{2}(t)
\end{equation}
and 
\begin{equation}
j_{k}(t)=\int_{0}^{t}J_{N}\left(\frac{2A_{k}(t')}{\omega}\right)dt'.
\end{equation}

\begin{figure}
 \includegraphics[height=4.5cm]{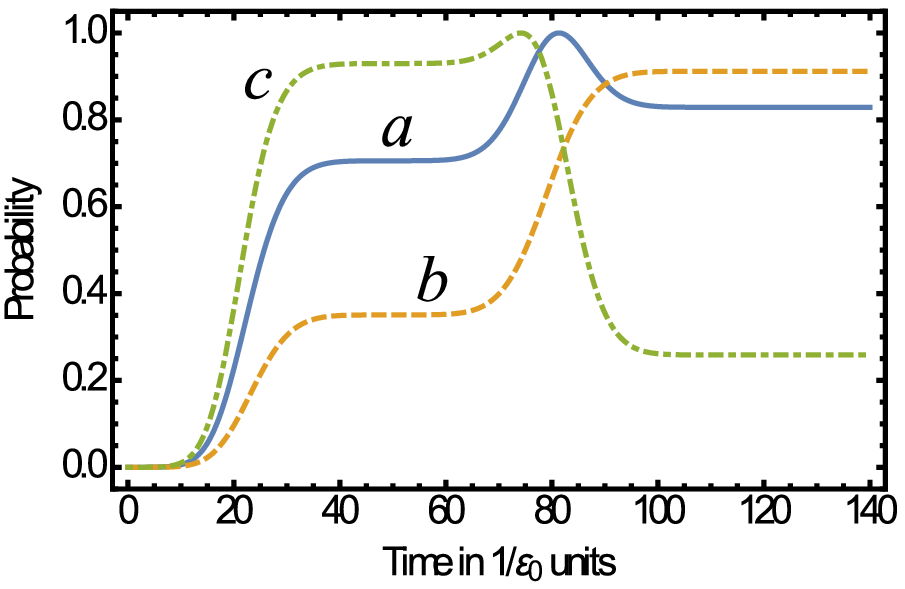}
\caption{\label{twoPulses1}Transition probability  population  $P_{2}(t)$  for the case of two-pulse excitation  in dependence of time-intervals  in units of $1/\epsilon_{0}$. The parameters are:  $\Delta/\varepsilon_{0} = 0.3$, $T\varepsilon_{0}=10$, $\tau/T=6$, $\omega/\varepsilon_{0} = 1$.  Solid line (a), $A_{0}/\varepsilon_{0}=0.19$;   dashed line (b),  $A_{0}/\varepsilon_{0}=0.12$;  dotdashed line (c), $A_{0}/\varepsilon_{0}=0.25$. The phase difference of two pulses is $\Delta\theta=\theta_{2}-\theta_{1}=0$.}
\end{figure}

\begin{figure}
 \includegraphics[height=4.5cm]{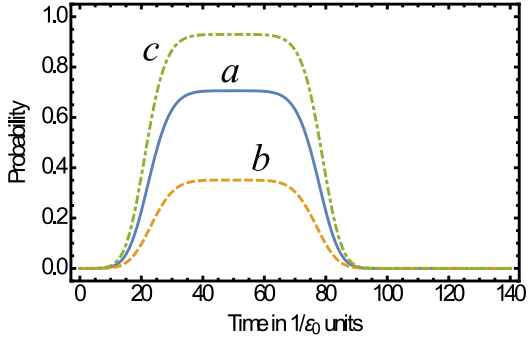}
\caption{\label{twoPulses2}Transition probability   $P_{2}(t)$  for the case of two-pulse excitation  in dependence of time-intervals  in units of $1/\epsilon_{0}$. The parameters are:  $\Delta/\varepsilon_{0} = 0.3$, $T\varepsilon_{0}=10$, $\tau/T=6$, $\omega/\varepsilon_{0} = 1$.  Solid line (a), $A_{0}/\varepsilon_{0}=0.19$;   dashed line (b),  $A_{0}/\varepsilon_{0}=0.12$;  dotdashed line (c), $A_{0}/\varepsilon_{0}=0.25$. The phase difference of two pulses is $\Delta\theta=\theta_{2}-\theta_{1}=\pi$.}
\end{figure}

This expression takes place for the general case of  multi-order resonant interactions in term of the factor $N$. As we see, in this configuration the controlling phase factor involved in the probability is the phase difference between consecutive pulses.

Now we focus on investigation of phase-dependent effects for one-photon resonant configuration considering pulses with Gaussian envelopes (\ref{GaussEnvelope}) that are shifted on
the time-interval $\tau$ between centers of the Gaussians. 

The results for the population of excited state for two cases of phase delay between pulses are depicted in the Figs. \ref{twoPulses1} and Figs. \ref{twoPulses2}. As we can see, high-level of the population $P_{2}(t)$ is realized for two-pulse excitation in dependence from the duration of pulses $T$ and time-interval between them $\tau$.
In Figs. \ref{twoPulses1}, if the envelope phase difference between two pulses is zero, $\Delta\theta=\theta_{2}-\theta_{1}=0$, we observe that  time-evolution of the population strongly depends from the amplitude of pulse field and the population inversion 
$P_{2}(t)-P_{1}(t)=2P_{2}(t)-1$ is realized for definite time-intervals. 
If the phase delay is changed to $\pi$, i. e. the difference between phases $\Delta\theta=\theta_{2}-\theta_{1}=\pi$,  the effect of the second pulse  reverse the
excitation process. The typical results are depicted in Figs. \ref{twoPulses2} for three values of the field amplitude. As we see, the curves indicate controllable excitation of qubit state by the first pulse and its decay produced by the second pulse. We conclude that these results depicted in  Figs. \ref{twoPulses1} and  Figs. \ref{twoPulses2} containing the interference effects between pulses are essentially differ from the analogous results for one-pulse regime (see,  Fig. \ref{singlePulse1}).

It is easy to realize from the formula (\ref{expExpand}) that dependence on the phase shift in probabilities varies with the order of the resonance due to the factor $\cos(N(\theta_{2}-\theta_{1}))$.  The order of the resonance reduces the required phase difference for observed effects by the order of resonance.  For example, in the case of second-order resonance the effect of probability raising and reducing analogous to the result shown on Fig. \ref{twoPulses2} takes place for $\Delta\theta=\pi/2$, instead of $\pi$, as it is for the case of first order resonance.

\subsection{The case of multi-qubit systems}

Recently,  systems composed
of several coupled qubits (see, e.g.,  \cite{Berkley}) are the subject of much attention. Therefore, in this subsection we shortly discuss  the modification of above approach for multi-qubit systems. The concrete calculations will be done for  two coupled qubits. The interaction of two-qubit system with driving field have been investigated in theoretical and experimental publications (see, e.g.,
 \cite{Majer,Steffen,Plantenberg,Groot,Shevchenko1,*Ilichev,AmpSpec, Ivanov}).

The effective Hamiltonian of multi-qubit system is written through the Pauli matrices and tunneling amplitudes of the individual qubits 
\begin{equation}
\hat{H}=\hat{H}_{0}-\sum_{i=1}^{N}\frac{\Delta_{a}}{2}\hat{\sigma}_{x}^{(a)},
\end{equation}
where
\begin{equation}
\hat{H}_{0}=-\sum_{a=1}^{N}g_{a}(t)\hat{\sigma}_{z}^{(a)}+\frac{1}{2}\sum_{a,b}J_{a,b}\hat{\sigma}_{z}^{(a)}\hat{\sigma}_{z}^{(b)},
\end{equation}
where time-dependent  functions $ g_{a}(t) $ are determined by Eq.(2) for individual qubits
\begin{equation}\label{SigmaZFactor}
g_{a}(t)=\frac{1}{2}\varepsilon_{a}+A_{a}(t)\cos(\omega_{a} t+\theta_{a})
\end{equation}
 and  $J_{a,b}$ are the coupling energies between qubits.

In this case the  system state in the Furry picture
$|\Psi(t)\rangle=\hat{R}(t)|\Phi(t)\rangle$  is calculated through the operator
\begin{equation}
\hat{R}(t)=\exp\bigg[i\sum_{a}\Lambda_{a}(t)\hat{\sigma}_{z}^{(a)}-\frac{it}{2}\sum_{a,b}J_{a,b}\hat{\sigma}_{z}^{(a)}\hat{\sigma}_{z}^{(b)}\bigg],
\end{equation}
where
\begin{equation}
\Lambda_{a}(t)=\int_{0}^{t}g_{a}(t')dt',
\end{equation}
in the following form
\begin{equation}
|\Phi(t)\rangle=T\exp\bigg(-i\int_{0}^{t}\hat{h}_{mq}(t')dt'\bigg)=\exp(\hat{\Omega}_{mq}(t))
\end{equation}
through the  Hamiltonian of multi-qubit system in Furry representation 
\begin{equation}
\hat{h}_{mq}(t)=-\frac{1}{2}\hat{R}^{-1}(t)\bigg(\sum_{a}\Delta_{a}\hat{\sigma}_{x}^{(a)}\bigg)\hat{R}(t)
\end{equation}
 and the corresponding Magnus operator $\hat{\Omega}_{mq}(t) $. Thus, the results of Magnus expantions (12), (13) take also  place for multi-qubit systems with the Hamiltonian  $\hat{h}_{mq}(t) $.

Below we calculate the Hamiltonian  $\hat{h}_{mq}(t) $ for a two-qubit system. In this case
\begin{equation}
\hat{H}=\hat{H}_{0}-\frac{1}{2}\left(\Delta_{1}\hat{\sigma}_{x}^{(1)}+\Delta_{2}\hat{\sigma}_{x}^{(2)}\right),
\end{equation}
where
\begin{equation}
\hat{H}_{0}=-g_{1}(t)\hat{\sigma}_{z}^{(1)}-g_{2}(t)\hat{\sigma}_{z}^{(2)}+J\hat{\sigma}_{z}^{(1)}\hat{\sigma}_{z}^{(2)}.
\end{equation}
Thus, the effective Hamiltonian of coupled two qubits in the Furry picture is written as
\begin{equation}\label{mqH}
\hat{h}_{mq}(t)=\hat{R}^{-1}(t)\left(\Delta_{1}\hat{\sigma}_{x}^{(1)}+\Delta_{2}\hat{\sigma}_{x}^{(2)}\right)\hat{R}(t),
\end{equation}
where
\begin{equation}
\hat{R}(t)=\exp\bigg[i\Lambda_{1}(t)\hat{\sigma}_{z}^{(1)}+i\Lambda_{2}(t)\hat{\sigma}_{z}^{(2)}-iJt\hat{\sigma}_{z}^{(1)}\hat{\sigma}_{z}^{(2)}\bigg]
\end{equation}
and $J=J_{1,2}=J_{2,1}$ is the strength of the interaction
between the qubits.

The transformed Hamiltonian is calculated in the following form 
\begin{widetext}
\begin{eqnarray}\label{twoQubitMH}
\hat{h}_{mq}(t)=\Delta_{1}\left(
\begin{array} {cccc}
0 & 0 & e^{-2i(Jt+\Lambda_{1}(t))} & 0\\
0 & 0 & 0 & e^{2i(Jt-\Lambda_{1}(t))}\\
e^{2i(Jt+\Lambda_{1}(t))} & 0 & 0 &0\\
0 & e^{-2i(Jt-\Lambda_{1}(t))} & 0 & 0 
\end{array}
\right)\nonumber\\
+\Delta_{2}\left(
\begin{array} {cccc}
0 & e^{-2i(Jt+\Lambda_{2}(t))} & 0 & 0\\
e^{2i(Jt+\Lambda_{2}(t))} & 0 & 0 &0\\
0 & 0 & 0 & e^{2i(Jt-\Lambda_{2}(t))}\\
0 & 0 & e^{-2i(Jt-\Lambda_{2}(t))} & 0 
\end{array}
\right)\nonumber\\
=\left(
\begin{array} {cccc}
0 & \Delta_{2}e^{-2i(Jt+\Lambda_{2}(t))} & \Delta_{1}e^{-2i(Jt+\Delta_{1}\Lambda_{1}(t))} & 0\\
\Delta_{2}e^{2i(Jt+\Lambda_{2}(t))} & 0 & 0 & \Delta_{1}e^{2i(Jt-\Lambda_{1}(t))}\\
\Delta_{1}e^{2i(Jt+\Lambda_{1}(t))} & 0 & 0 & \Delta_{2}e^{2i(Jt-\Lambda_{2}(t))}\\
0 & \Delta_{1}e^{-2i(Jt-\Lambda_{1}(t))} & \Delta_{2}e^{-2i(Jt-\Lambda_{2}(t))} & 0 
\end{array}
\right)
\end{eqnarray}
\end{widetext}
and the  details of calculations are presented in the Appendix. This Hamiltonian operates on the four-level system of coupled qubits.  
It should be mentioned that this result  is obtained in the general form and describes  two-qubit system interacting with two component driving field with arbitrary time-dependent amplitudes and phases.

\section{Multiple resonant excitations with sequential pulses}\label{MRESP}

We focus now on the interaction of the qubit with a train of pulses. A more general case is when the external field is a sum of  pulses with the envelopes $A_{k}(t)$ and with the phases $\theta_{k}$, thus
\begin{equation}
g(t)=\frac{1}{2}\varepsilon_{0}+\sum_{k=1}^{M}A_{k}(t)\cos(\omega t+\theta_{k}). 
\end{equation}
Below we analyze the effect of a sequence of pulses considering the case of not overlapped envelopes $A_{k}(t)$, which means independent qubit-pulse  interactions for every consecutive pulse. The overall effect of pulses in this configuration depends on phase changes during the pauses.  In this case the exponential function reads as
\begin{equation}
e^{2i\Lambda(t)}=(-1)^{N}\sum_{k}^{M}J_{N}\left(\frac{2A_{k}(t)}{\omega}\right)e^{-iN\theta_{k}},
\end{equation}
while the components of the vector $\vec{\rho}(t)=\vec{G}(t)/|\vec{G}(t)|$ from Eq. (\ref{phaseGComp}) can be calculated in the following form
\begin{eqnarray}\label{phaseGCompMulti}
G_{x}(t) & \approx & (-1)^{N}\Delta\sum_{k=1}^{M}\cos(N\theta_{k})\int_{0}^{t}J_{N}\left(\frac{2A_{k}(t')}{\omega}\right)dt',\nonumber\\
G_{y}(t) & \approx & (-1)^{N}\Delta\sum_{k=1}^{M}\sin(N\theta_{k})\int_{0}^{t}J_{N}\left(\frac{2A_{k}(t')}{\omega}\right)dt',\\
G_{z}(t) & \approx & 0.\nonumber
\end{eqnarray}

On the whole the probability of qubit state excitation reads as
\begin{equation}
P_{2}(t)=\sin^{2}\left(G(t)\right).
\end{equation}

The obtained results are valid for an arbitrary multi-resonant interaction, where the order of resonance $N$ is indicated by the index of the Bessel function and by phase-argument in sine/cosine (see formula (\ref{phaseGCompMulti})). These results are applicable for an important case naturally realized in practice that is the train consisting of identical pulses with equal pulse areas and different phases that can be obtained by using various modulators.

\subsection{Excitation by  identical pulses: formation of quasienergetic states}

Another interesting configuration that can be realized involves the train consisting of identical pulse envelopes with the same phases. The trains consisting of big numbers of pulses approximately display periodicity in time that can be exploited for realization of new regimes of qubit dynamics.

For a sequence of pulses with identical $\theta_{k}$ phases $G(t)$ can be written as
\begin{multline}
G(t)=\Delta\sum_{k=1}^{M}\int_{0}^{t}J_{N}\left(\frac{2A_{k}(t')}{\omega}\right)dt'\\
=\Delta\int_{0}^{t}J_{N}\left(\frac{2A(t')}{\omega}\right)dt',
\end{multline}
where
\begin{equation}
A(t)=\sum_{k=1}^{M}A_{k}(t).
\end{equation}
This formula is obtained in approximation of not overlapping envelopes, where interactions conditioned by pulses are separated by interaction-free time intervals. Thus, we get to the equation
\begin{equation}
G(t)=\Delta\int_{0}^{t}J_{N}\left(\frac{2A(t')}{\omega}\right)dt'.
\end{equation}

To evaluate these quantities further we need to make  integration of a periodic function $J_{N}\left(\frac{2A(t')}{\omega}\right)$ with the period that equals to time intervals between pulses. For train of Gaussian pulses
\begin{equation}\label{A}
A(t)=A_{0}\sum_{k=-\infty}^{\infty}e^{-\frac{(t-k\tau)^{2}}{T^{2}}}
\end{equation}
the period equals to $\tau$.

It is easy to represent the function $G(t)$ as the sum of  term with linear dependence of time and periodic on time term
\begin{equation}
G(t)=\Delta\gamma_{N}t+\phi_{N}(t).
\end{equation}
Here, $\phi_{N}(t)$ is a periodic function with period $\tau$ defined in $t\in[t_{0},t_{0}+\tau]$ as
\begin{equation}
\phi_{N}(t)=\Delta\int_{t_{0}}^{t}\left(J_{N}\left(\frac{2A(t')}{\omega}\right)-\gamma_{N}\right)dt'
\end{equation}
and 
\begin{equation}
\gamma_{N}=\frac{1}{\tau}\int_{t_{0}}^{t_{0}+\tau}J_{N}\left(\frac{2A(t')}{\omega}\right)dt'
\end{equation}
is the Bessel function averaged on the period.
It is easy to realize that the last expression can be written through the integral of Bessel function for just one of the identical pulse envelopes as

\begin{figure}
 \includegraphics[height=4.5cm]{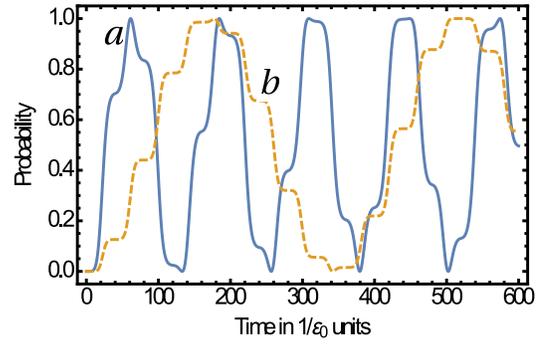}
\caption{\label{multipulseCase} Transition probability   $P_{2}(t)$  for the case of identical pulses train in dependence of time-intervals, in units of $1/\epsilon_{0}$. The parameters are:  $\Delta/\varepsilon_{0} = 0.3$, $T\varepsilon_{0}=10$, $\omega/\varepsilon_{0} = 1$. Solid line (a) describes  resonance of the first order, with $A_{0}/\varepsilon_{0}=0.19$, $\tau/T=4$;  dashed line (b) describes resonance of the second order, with $A_{0}/\varepsilon_{0}=0.5$.}
\end{figure}

\begin{figure}
 \includegraphics[height=4.5cm]{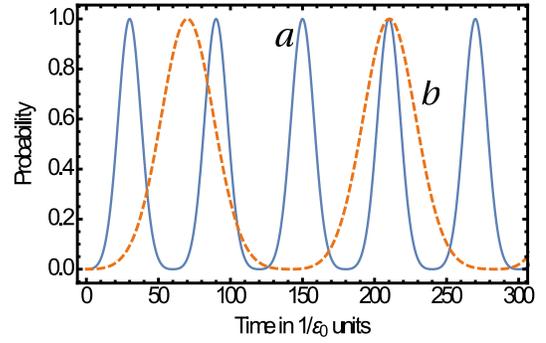}
\caption{\label{multipulseCasePeriodic}Population of excited state $P_{2}(t)$ of the qubit under sequence of identical pulses showing regular behaviour in the case of $E_{N}\tau=\pi$. Solid line (a) describes resonance of the first order, $\Delta/\varepsilon_{0}=0.3$, $T\varepsilon_{0}=20$,  $\tau/T=3$ and $A_{0}/\varepsilon_{0}=0.315$;  dashed line (b) describes  resonance of the second order, $\Delta/\varepsilon_{0}=0.4$, $T\varepsilon_{0}=65$,  $\tau/T=2.15$ and $A_{0}/\varepsilon_{0}=0.457$.}
\end{figure}

\begin{equation}
\gamma_{N}=\frac{1}{\tau}\int_{-\infty}^{\infty}J_{N}\left(\frac{2A_{k}(t)}{\omega}\right)dt.
\end{equation}

The population of the excited states (if the system was initially in the lower state, $C_{1}(0) = 1$, $C_{2}(0) = 0$ ) as a function of time is then given by
\begin{equation}\label{P2IdenticalPulses}
P_{2}(t)=\sin^{2}\left(\Delta\gamma_{N}t+\phi_{N}(t)\right).
\end{equation}

It should be mentioned that the formulae (\ref{A}) allows to introduce the quasienergetic states and quasienergies for the qubit interacting with  time-periodic sequence of pulses.  Indeed, it is easy to check that the solution of Eq. (\ref{stateEqMag}) for this configuration  can be expressed in the adiabatic basis
\begin{equation}
|\pm\rangle=\frac{1}{\sqrt{2}}\left(|1\rangle\pm|2\rangle\right)
\end{equation}
 as the quasienergetic states
\begin{equation}\label{quasienergy_rep}
|\Phi_{N}^{\pm}\rangle=e^{\pm iG(t)}|\pm\rangle=e^{\pm i(E_{N}t+\phi_{N}(t))}|\pm\rangle
\end{equation}
with the quasienergies 
\begin{equation}\label{quasienergies}
E_{N}^{+}=-E_{N}^{-}=E_{N}=\frac{\Delta}{\tau}\int_{-\infty}^{\infty}J_{N}\left(\frac{2A_{k}(t)}{\omega}\right)dt.
\end{equation}
The Eq. (\ref{quasienergy_rep}) can be written also as
\begin{equation}
|\Phi_{N}^{\pm}\rangle=e^{\mp iE_{N}t}U_{N}^{\pm}(t)|\pm\rangle,
\end{equation}
where
\begin{equation}
U_{N}^{\pm}(t)=e^{\mp i\phi_{N}(t)}
\end{equation}
is periodic on time in accordance with denotation of quasienergetic states
\begin{equation}
U_{N}^{\pm}(t+\tau)=U_{N}^{\pm}(t).
\end{equation}
It is important also to rewrite quasienergies with the input of qubit energy levels. In this way, taking into account the relation
\begin{equation}
|\Psi^{\pm}_{N}(t)\rangle=e^{i\Lambda(t)\hat{\sigma}_{z}}|\Phi^{\pm}_{N}(t)\rangle,
\end{equation}
we get the states of driven qubit in the following form
\begin{subequations}
\begin{eqnarray}
|\Psi^{+}_{N}(t)\rangle & = & e^{-i\left(\varepsilon_{1}+E^{+}_{N}\right)t}U^{+}_{N}(t)\bigg[e^{-i\sum_{k}A_{k}(t)\cos(\omega t + \theta_{k})}|1\rangle\nonumber\\
& + & e^{i\varepsilon_{0}t}e^{i\sum_{k}A_{k}(t)\cos(\omega t + \theta_{k})}|2\rangle\bigg],\\
|\Psi^{-}_{N}(t)\rangle & = & e^{-i\left(\varepsilon_{2}+E^{-}_{N}\right)t}U^{-}_{N}(t)\bigg[e^{i\varepsilon_{0}t}e^{-i\sum_{k}A_{k}(t)\cos(\omega t + \theta_{k})}|1\rangle\nonumber\\
& - & e^{i\sum_{k}A_{k}(t)\cos(\omega t + \theta_{k})}|2\rangle\bigg].
\end{eqnarray}
\end{subequations}
In these wave functions we added  the phase factors involving the  sum of qubit energetic levels, that is zero for the case of the truncated Hamiltonian (\ref{Ham1}), in which the half of the sum of qubit energetic levels has been omitted.

From the last equations the relation between real energies and quasienergies can be written as
\begin{subequations}
\begin{eqnarray}
E_{1,N}=\varepsilon_{1}+E^{+}_{N}=\varepsilon_{1}+E_{N},\\
E_{2,N}=\varepsilon_{2}+E^{-}_{N}=\varepsilon_{2}-E_{N}.
\end{eqnarray}
\end{subequations}
 Note, that the sum of two quasienergies obeys the relation $E^{+}_{N}+E^{-}_{N}=0$, while $E_{1,N}+E_{2,N}=\varepsilon_{1}+\varepsilon_{2}$. This result is in accordance with the exact result taking place for a two-level atom in a monochromatic field \cite{Shirley}.

\begin{figure}
 \includegraphics[height=4.5cm]{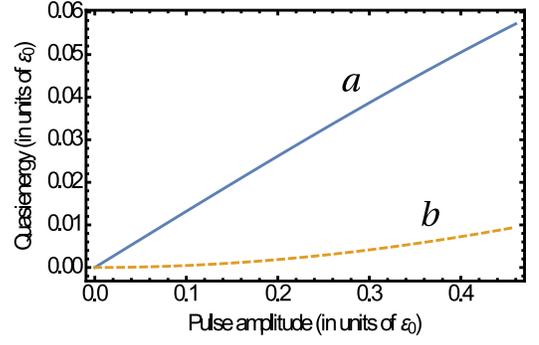}
\caption{\label{quasiEnGr} Quasienergy  $E_{N}$  depending on the  pulse amplitude  in units of $\varepsilon_{0}$. The parameters are: $\Delta/\varepsilon_{0}=0.3$,  the pulse width $T\varepsilon_{0}=10$, $\tau/T=4$, $\omega/\varepsilon_{0}=1$.  Solid line (a) describes  the first-order resonant configuration;  dashed line (b) describes the second order resonance.}
\end{figure}

\begin{figure}
 \includegraphics[height=4.5cm]{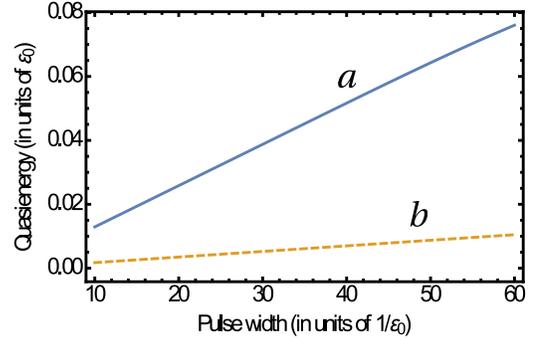}
\caption{\label{quasiEnGrWidth} Quasienergy $E_{N}$ depending on pulse duration. The parameters are:  $\Delta/\varepsilon_{0}=0.4$, $\omega/\varepsilon_{0}=1$, $\tau\varepsilon_{0}=40$, the  amplitude $A_{0}/\varepsilon_{0}=0.38$. Solid line (a) and dashed line (b) describe the cases of first- and second-order  resonances, respectively.}
\end{figure}

Now we analyse time-evolution of the  probability of excitation (\ref{P2IdenticalPulses}). The typical results are depicted in Fig. (\ref{multipulseCase}) for two cases of first-order and second-order resonances. As we see the populations for both  resonance configurations show complete population inversion for definite time-intervals.   Such result is cardinally different from the case of single-pulse excitation, where the population  $P_{2}(t)$  for the second-order resonance is small. It can be observed from comparison of  Fig. (\ref{singlePulse1}) and Fig. (\ref{singlePulse2}). 
The function $G(t)=\Delta\gamma_{N}t+\phi_{N}(t)$  is an increasing function in time but it grows also periodically due to its ``linear+periodic'' structure. Therefore, the dynamics of populations Eq. (\ref{P2IdenticalPulses}) seems to be aperiodic in time as is demonstrated in  Fig. \ref{multipulseCase}.

An interesting regime of Rabi oscillations can be  formed if the following condition for the quasienergy  takes place
\begin{equation}
\Delta\gamma_{N}\tau=E_{N}\tau=\pi m
\end{equation} 
or in the following form
\begin{equation}
\Delta\int_{-\infty}^{\infty}J_{N}\left(\frac{2A_{k}(t)}{\omega}\right)dt=\pi m,
\end{equation}
where $m=1,2,...$.
In this case we observe that $G(t+\tau)=\Delta\gamma_{N}t+\phi_{N}(t)+\pi m=G(t)+\pi m$ and hence the transition probability $ P_{2}(t)=\sin^{2}(G(t))$  shows periodicity on time-intervals.  Thus, for the definite values of quasienergy the transition probability displays regular, periodic dynamics. The typical results showing clear regular behaviour are depicted in  Fig. \ref{multipulseCasePeriodic}.

In the end of this Sec. we analyse  the quasienergy in dependence from the amplitude of pulses and their duration. Calculations  on base of Eq. (\ref{quasienergies})  lead to  large values  of the quasienergy in comparison with the analogous results for two-level atomic system.   As we see in   Fig. \ref{quasiEnGr} and  Fig. \ref{quasiEnGrWidth} the quasienergy increases almost  linearly with increasing the amplitude of pulses as well as with the duration of pulses for both resonant configurations.

\section{Conclusion}\label{conclusion}

In conclusion, we have investigated the dynamics and manipulations of qubit by its multiple resonant interactions with external pulses. Single qubit  quantum operations have been proposed and analyzed for various regimes of state excitation including: single-pulse excitation, two-pulse excitation with phase shift between pulse envelopes being controlling parameter and for excitation with identical sequential pulses. This investigation have been performed in the approach,  based on Bloch vector's treatment and the Magnus quantum electrodynamical formalism in Furry picture that can be applied for various types of interactions. 

This work has showed the possibility to produce a large variety of qubit operations for time-domain in dependence of the amplitude of driven field, the parameters of pulses, the tunneling rate and the  controlling phase factors.  Considering two resonant configurations, i.e. first- and second-order resonant configurations,  we have shown that the complete population inversion can be achieved through tailoring the pulse train for both configurations at definite time intervals.
 
Particular emphasis has been laid on interaction of qubit with train of identical pulses. We have shown that quasienergies and quasienergetic states of combined system "qubit + pulse train"  are forming  for all multiple resonant regimes. The quasienergy  increases almost  linearly with increasing the amplitude of pulses as well as with the duration of pulses.  Moreover, we have demonstrated that for all multiple resonant regimes the  state population exhibits time-dependent aperiodic oscillations, but also displays   periodically regular oscillations for the definite values of the quasienergy, $E_{N}\tau=\pi m$.

This approach has been formulated  in general case for analysing the Rabi model with time-dependent  interaction Hamiltonians along  the $z$ axis for multi-resonant interactions. Nevertheless, we demonstrate that the approach proposed can be used for pulse-driven multi-qubit systems  as well as for  interaction of qubits beyond RWA.  Thus,  an interesting question to address would be application of the general expressions obtained in this paper for qubit operations beyond RWA and for elaboration of quantum gates with 
several qubits.

\begin{acknowledgments}
We acknowledge support from the Armenian State Committee of Science, the Project No.13-1C031.
\end{acknowledgments}

\appendix*

\section{}

To calculate the Hamiltonian (\ref{mqH}) we use the following well-known formulas 

\begin{subequations}
\begin{eqnarray}
e^{i\alpha\hat{\sigma}_{z}}\hat{\sigma}_{x}e^{-i\alpha\hat{\sigma}_{z}}=\cos(2\alpha)\hat{\sigma}_{x}-\sin(2\alpha)\hat{\sigma}_{y},\\
e^{i\alpha\hat{\sigma}_{z}}\hat{\sigma}_{y}e^{-i\alpha\hat{\sigma}_{z}}=\sin(2\alpha)\hat{\sigma}_{x}+\cos(2\alpha)\hat{\sigma}_{y},
\end{eqnarray}
\end{subequations}
 that allows us to performe unitary Furry transfomations. In this way, for the Pauli matrice one of the qubit we obtain the  following  operator equation
\begin{multline}
\hat{R}^{-1}(t)\hat{\sigma}_{x}^{(1)}\hat{R}(t)\\
=\cos(2\Lambda_{1}(t))\left(\hat{\sigma}_{x}^{(1)}\cos(2Jt\hat{\sigma}_{z}^{(2)})-\hat{\sigma}_{y}^{(1)}\sin(2Jt\hat{\sigma}_{z}^{(2)})\right)\\
-\sin(2\Lambda_{1}(t))\left(\hat{\sigma}_{x}^{(1)}\sin(2Jt\hat{\sigma}_{z}^{(2)})+\hat{\sigma}_{y}^{(1)}\cos(2Jt\hat{\sigma}_{z}^{(2)})\right).
\end{multline}
Then, this equation is transfomed in the matrice form as the following
\begin{eqnarray}
\hat{R}^{-1}(t)\hat{\sigma}_{x}^{1}\hat{R}(t)=\left(
\begin{array} {cc}
0 & e^{-2iJt\hat{\sigma}_{z}^{(2)}}e^{-2i\Lambda_{1}(t)}\\
e^{2iJt\hat{\sigma}_{z}^{(2)}}e^{2i\Lambda_{1}(t)} & 0
\end{array}
\right)\nonumber\\
=\left(
\begin{array} {cccc}
0 & 0 & e^{-2i(Jt+\Lambda_{1}(t))} & 0\\
0 & 0 & 0 & e^{2i(Jt-\Lambda_{1}(t))}\\
e^{2i(Jt+\Lambda_{1}(t))} & 0 & 0 &0\\
0 & e^{-2i(Jt-\Lambda_{1}(t))} & 0 & 0 
\end{array}
\right)
\end{eqnarray}
The expression $\hat{R}^{-1}(t)\hat{\sigma}_{x}^{(2)}\hat{R}(t)$ is evaluated in a similar way. On the whole, these results lead to the formula (\ref{twoQubitMH}).

\nocite{*}

\bibliographystyle{apsrev4-1}
\bibliography{abovyan}

\end{document}